\documentclass[a4paper]{article}

\usepackage{graphicx}
\usepackage{mathtools}
\usepackage{multirow}
\usepackage[utf8]{inputenc}
\usepackage{amsfonts}
\usepackage[margin=1.2in]{geometry}

\providecommand{\keywords}[1]
{
  \small	
  \textbf{\textit{Keywords---}} #1
}

\usepackage[sorting=none]{biblatex}
\usepackage[colorlinks]{hyperref}
\begin{document}

\title{Stress and Displacement Fields in the Surface of an Elastic Half-Space under the Action of a Constant Shear Load over a Circular Contact Domain}

\author{E. Willert \\
Technische Universit\"at Berlin\\
Stra\ss{}e des 17. Juni 135, 10623 Berlin, Germany \\
e.willert@tu-berlin.de}
\maketitle

\begin{abstract}
Based on the superposition of incremental frictional surface tractions that, in the case of an incompressible elastic half-space, correspond to a rigid tangential translation of a circular contact domain, the stress and displacement fields in the surface are determined in closed analytic form for an elastic half-space under the action of a constant shear traction over a circular contact domain on the surface. The obtained exact solutions can serve as benchmarks for numerical models, or can be used for the deeper contact-mechanical analysis of adhesive sliding contacts of soft materials.
\end{abstract}


\section{Introduction}\label{sec1} 

The interplay between friction and adhesion has attracted a lot of scientific interest in the past decade. Especially for biomaterials, or other soft media, adhesion plays an important role. In sliding contacts of such materials, it is often found that the frictional stress distribution in the (real) adhesive contact area is (approximately) constant -- indepedent of the normal load and the respective contact area -- corresponding to a critical shear stress $\tau_0$ to initiate the sliding mode (see \cite{Mengaetal2018}, \cite{Liashenkoetal2024}, and the references therein).

Thus, one aspect of the interaction between adhesion and friction, which can be studied quite easily within the framework of such a "non-Coulomb" type friction law, is the elastic coupling, e.g., the influence of the frictional tractions on the normal contact solution and the respective contact area. As the shear tractions are constant, independent of the normal load, and, from a contact-mechanical perspective, known (as an interface property), this approach can be seen as a "counterpart" of the Goodman approximation \cite{Goodman1962} of elastic coupling -- where the normal contact solution is assumed to be known (from the respective uncoupled problem), and only its influence on the tangential contact problem is considered (and not vice versa).

To apply this spproach, we first require the full contact solution (including stress and displacement fields, at least, in the surface) for the action of a constant shear traction. In the case of a circular contact domain on the surface of an homogeneous, isotropic, linearly elastic half-space, some aspects of this contact problem have already been considered in \cite{Mengaetal2018}. The present manuscript extends that solution to the complete stress and displacement fields in the surface. While the authors in \cite{Mengaetal2018} use field-point integration of the fundamental solution to the Cerruti problem \cite{Cerruti1882}, in the following, a more elegant approach is used, based on the superposition of fields under the action of a specific tangential surface traction over a circular contact domain, which were published recently in exact form by the author \cite{Willert2024}. 

\section{Problem Statement}\label{sec2}

Let us consider an homogeneous, isotropic, linearly elastic half-space with shear modulus $G$ and Poisson ratio $\nu$, under the action of a constant shear traction $\tau_0$ over a circular domain with radius $a$ at the surface of the half-space. Let the center of that domain be the origin of a cartesian coordinate system $\left\lbrace x, y, z\right\rbrace$, with $z$ pointing into the half-space, and $x$ into the direction of the shear traction. We seek the stress and displacement fields in the half-space surface.

\section{"Flat Punch" Solution}\label{sec3}

We will determine these fields via the superposition of fields under the action of a shear traction distribution
\begin{equation}
\sigma_{xz}\left(r,z = 0\right) = \frac{4G}{\pi\left(2 - \nu \right)}\frac{u_0}{\sqrt{\xi^2 - r^2}} \quad , \quad r < \xi, \label{eq_FP_traction}
\end{equation}
with the polar radius $r = \sqrt{x^2 + y^2}$ and constants $u_0$ and $\xi$, which for an incompressible elastic half-space corresponds to a rigid tangential displacement of a circular contact domain with radius $\xi$ by $u_0$. The stress and displacements fields in and below the surface for these tractions have been published recently in exact form by the author \cite{Willert2024}. In the following, we will only consider the values in the surface $z = 0$. However, the subsurface fields could also be determined for the constant shear over a circular contact domain, in complete analogy to the considerations in Sect. \ref{sec4}.

\subsection{Surface Displacements}

First, let us consider the surface displacements under the action of the traction distribution \eqref{eq_FP_traction}. These have not been given explicitly in \cite{Willert2024}, but can be calculated easily from the general soluton for the subsurface displacement field given in that work.

For the displacements in the surface, but outside the contact domain ($\xi < r$), we have
\begin{eqnarray}
\frac{\pi \left(2 - \nu\right)}{2u_0}~u_x &=& \left(2 - \nu \right)\arcsin\left(\frac{\xi}{r}\right) + \nu ~ \frac{x^2 - y^2}{r^2}\frac{\xi \sqrt{r^2 - \xi^2}}{r^2}, \nonumber \\
\frac{\pi \left(2 - \nu\right)}{2u_0}~u_y &=& 2\nu ~ \frac{xy}{r^2}\frac{\xi \sqrt{r^2 - \xi^2}}{r^2}, \\
\frac{\pi \left(2 - \nu\right)}{2u_0}~u_z &=& \frac{x\left(1 - 2\nu \right)\xi }{r^2}. \nonumber
\end{eqnarray}
For the contact domain ($r \le \xi$), the non-vanishing displacement components are
\begin{eqnarray}
u_x &=& u_0 \nonumber \\
\frac{\pi \left(2 - \nu\right)}{2u_0}u_z &=& x\left(1 - 2\nu \right)\left(\frac{\xi}{r^2} - \frac{\sqrt{\xi^2 - r^2}}{r^2}\right).
\end{eqnarray}

\subsection{Surface Stresses}

In the surface, inside the contact domain, the only non-vanishing physical stress component is given by the load \eqref{eq_FP_traction}.

In the surface, but outside the contact domain, the non-vanishing physical stress components are \cite{Willert2024}
\begin{eqnarray}
\frac{\pi \left(2 - \nu \right)}{4G u_0}~\sigma_{xy} &=& -\frac{y \xi}{r^4}\left[ \frac{r^2}{\sqrt{r^2-\xi^2}} - 2\nu\left\lbrace \left(1 - \frac{4x^2}{r^2}\right)\sqrt{r^2-\xi^2} + \frac{x^2}{\sqrt{r^2-\xi^2}}\right\rbrace \right], \nonumber \\
\frac{\pi \left(2 - \nu \right)}{4G u_0}~\sigma_{xx} &=& -\frac{x \xi}{r^4}\left[ \frac{2r^2}{\sqrt{r^2-\xi^2}} - 2\nu\left\lbrace \left(3 - \frac{4x^2}{r^2}\right)\sqrt{r^2-\xi^2} - \frac{y^2}{\sqrt{r^2-\xi^2}}\right\rbrace \right], \\
\frac{\pi \left(2 - \nu \right)}{4G u_0}~\sigma_{yy} &=& -2\nu ~ \frac{x \xi}{r^4}\left[\left(3 - \frac{4x^2}{r^2}\right)\sqrt{r^2-\xi^2} + \frac{x^2}{\sqrt{r^2-\xi^2}} \right], \nonumber
\end{eqnarray}

\section{Solution for Constant Shear over a Circular Domain}\label{sec4}

Let us now turn our attention back to the original problem of interest, the constant shear load over a circular contact domain. One way of solving this problem would be field-point integration of the fundamental solution to the Cerruti problem of a tangential point force at the surface of the half-space. A far more elegant procedure is the superposition of solutions for incremental loadings of the form \eqref{eq_FP_traction} with incremental "rigid displacements" $\text{d}u_0\left(\xi \right)$ and increasing radii $\xi$.

\subsection{"Flat Punch" Superposition}

The final shear stress distribution at the end of these incremental tangential loading steps must be given by the constant distribution over a circular domain with radius $a$. Hence,
\begin{equation}
\tau_0 \equiv \frac{4G}{\pi \left(2 - \nu \right)}\int_r^a \frac{u_0'\left(\xi \right) \text{d}\xi}{\sqrt{\xi^2 - r^2}} \quad , \quad r \le a.
\end{equation}
This is an Abel-type integral transform, which can be inverted with the result
\begin{equation}
u_0'\left(\xi \right) = \frac{\tau_0 \left(2 - \nu \right)}{2G}\frac{\xi}{\sqrt{a^2 - \xi^2}} \quad , \quad \xi \le a. \label{eq_series}
\end{equation}

The last equation gives the series of incremental "flat punch" loadings necessary to produce a constant shear stress distribution over a circular contact domain. Due to the linearity of all underlying equations, the stress and displacement fields arising from such a constant shear traction are given by the fields arising from the series of incremental loadings characterized by Eq. \eqref{eq_series}, while the fields for a single of these incremental loading steps have been given in Sect. \ref{sec3}.

\subsection{Surface Displacements}

\subsubsection{Displacements inside the Contact Area}

For the tangential displacements inside the contact area ($r \le a$), we have to distinguish two phases of the incremental loading (with increasing radii $\xi$): while $\xi < r$, the considered point at $r$ lies outside the "flat punch" contact domain; for $r \le \xi \le a$, it lies inside. Hence,

\begin{eqnarray}
\frac{\pi G u_x}{\tau_0} &=&  \frac{\pi}{2}\left(2 - \nu \right) \int_r^a \frac{\xi ~ \text{d}\xi}{\sqrt{a^2 - \xi^2}} + \left(2 - \nu \right)\int_0^r \arcsin\left(\frac{\xi}{r}\right)\frac{\xi ~ \text{d}\xi}{\sqrt{a^2 - \xi^2}} + \nu ~ \frac{x^2 - y^2}{r^4}\int_0^r \frac{\xi^2 \sqrt{r^2 - \xi^2} ~ \text{d}\xi}{\sqrt{a^2 - \xi^2}} \nonumber \\
&=&  \left(2 - \nu\right)a \text{E}\left(\rho \right) + \nu a^3 \frac{x^2 - y^2}{3r^4}\left[2\left(\text{E}\left(\rho \right) - \left(1 - \rho^2 \right)\text{K}\left(\rho \right)\right) - \rho^2 \text{E}\left(\rho \right)\right], \quad \rho = r / a,
\end{eqnarray}
which was already determined in \cite{Mengaetal2018} via field-point integration, with the Legendre form of the complete elliptic integrals of the first and second kind,
\begin{eqnarray}
\text{K}(k) &=& \int_0^{\pi/2}\left(1 - k^2 \sin^2 \varphi \right)^{-1/2}\text{d}\varphi,\nonumber \\
\text{E}(k) &=& \int_0^{\pi/2}\left(1 - k^2 \sin^2 \varphi \right)^{1/2}d\varphi.
\end{eqnarray}

Similarly, for the lateral displacements,
\begin{eqnarray}
\frac{\pi G u_y}{\tau_0} &=& 2\nu ~ \frac{xy}{r^4}\int_0^r \frac{\xi^2 \sqrt{r^2 - \xi^2} ~ \text{d}\xi}{\sqrt{a^2 - \xi^2}} \nonumber \\
&=& 2\nu a^3 \frac{xy}{3r^4}\left[2\left(\text{E}\left(\rho \right) - \left(1 - \rho^2 \right)\text{K}\left(\rho \right)\right) - \rho^2 \text{E}\left(\rho \right)\right],
\end{eqnarray}
and the normal displacements
\begin{eqnarray}
\frac{\pi G u_z}{\tau_0} &=& x\left(1 - 2\nu \right)\left\lbrace \int_0^a \frac{\xi^2 ~ \text{d}\xi}{r^2 \sqrt{a^2 - \xi^2}} - \int_r^a \frac{\xi \sqrt{\xi^2 - r^2} ~ \text{d}\xi}{r^2 \sqrt{a^2 - \xi^2}}\right\rbrace \nonumber \\
&=& \frac{\pi x \left(1 - 2\nu \right)}{4}.
\end{eqnarray}

Interestingly, the normal displacements correspond to a tilt of the contact area around the lateral axis by the (small) angle $\tau_0 (1 - 2\nu) / (4G)$.

\subsubsection{Displacements outside the Contact Area}

The points outside the contact area are, of course, outside all loading circles with increasing radii $\xi$. Hence,

\begin{eqnarray}
\frac{\pi G u_x}{\tau_0} &=& \left(2 - \nu \right) \int_0^a \arcsin \left(\frac{\xi}{r}\right) \frac{\xi ~ \text{d}\xi}{\sqrt{a^2 - \xi^2}} + \nu ~ \frac{x^2 - y^2}{r^4} \int_0^a \frac{\xi^2 \sqrt{r^2 - \xi^2} ~\text{d}\xi}{\sqrt{a^2 - \xi^2}} \nonumber \\
&=& \left(2 - \nu \right) \left[r\text{E}\left(\frac{1}{\rho }\right) + \frac{a^2}{r}\left(1 - \rho^2 \right)\text{K}\left(\frac{1}{\rho }\right)\right] + \\
&& \quad + \nu a^2 \frac{x^2 - y^2}{r^3}\left[\text{E}\left(\frac{1}{\rho }\right) - \frac{1}{3}\left(\left(1 + \rho^2\right)\text{E}\left(\frac{1}{\rho }\right) + \left(1 - \rho^2\right)\text{K}\left(\frac{1}{\rho }\right)\right)\right], \nonumber
\end{eqnarray}
and
\begin{eqnarray}
\frac{\pi G u_y}{\tau_0} &=& 2\nu ~ \frac{xy}{r^4}\int_0^a \frac{\xi^2 \sqrt{r^2 - \xi^2} ~ \text{d}\xi}{\sqrt{a^2 - \xi^2}} \nonumber \\
&=& 2\nu a^2 \frac{xy}{r^3}\left[\text{E}\left(\frac{1}{\rho }\right) - \frac{1}{3}\left(\left(1 + \rho^2\right)\text{E}\left(\frac{1}{\rho }\right) + \left(1 - \rho^2\right)\text{K}\left(\frac{1}{\rho }\right)\right)\right],
\end{eqnarray}
and
\begin{eqnarray}
\frac{\pi G u_z}{\tau_0} &=& x\left(1 - 2\nu \right) \int_0^a \frac{\xi^2 ~ \text{d}\xi}{r^2 \sqrt{a^2 - \xi^2}} = \frac{\pi x \left(1 - 2\nu \right)a^2}{4r^2}.
\end{eqnarray}

\subsection{Surface Stresses}

In complete analogy to the surface displacements, the non-trivial components of the surface stress field can be determined exactly, based on the aforementioned superposition idea. We have:

\subsubsection{Stresses inside the Contact Area}

\begin{eqnarray}
\frac{\pi \sigma_{xy}}{2\tau_0} &=& -\frac{y}{r^2}\left(1 - 2\nu ~ \frac{x^2}{r^2}\right)\int_0^r \frac{\xi^2 ~ \text{d}\xi}{\sqrt{r^2 - \xi^2}\sqrt{a^2 - \xi^2}} + 2\nu ~ \frac{y}{r^4}\left(1 - \frac{4x^2}{r^2}\right) \int_0^r \frac{\xi^2 \sqrt{r^2 - \xi^2} ~ \text{d}\xi}{\sqrt{a^2 - \xi^2}} \\
&=& -\frac{ya}{r^2}\left(1 - 2\nu ~ \frac{x^2}{r^2}\right)\left[\text{K}\left(\rho\right) - \text{E}\left(\rho \right)\right] + 2\nu ~ \frac{ya^3}{3r^4}\left(1 - \frac{4x^2}{r^2}\right)\left[2\left(\text{E}\left(\rho \right) - \left(1 - \rho^2 \right)\text{K}\left(\rho \right)\right) - \rho^2 \text{E}\left(\rho \right)\right],\nonumber
\end{eqnarray}
\begin{eqnarray}
\frac{\pi \sigma_{xx}}{2\tau_0} &=& -\frac{2x}{r^2}\left(1 + \nu ~ \frac{y^2}{r^2}\right)\int_0^r \frac{\xi^2 ~ \text{d}\xi}{\sqrt{r^2 - \xi^2}\sqrt{a^2 - \xi^2}} + 2\nu ~ \frac{x}{r^4}\left(3 - \frac{4x^2}{r^2}\right) \int_0^r \frac{\xi^2 \sqrt{r^2 - \xi^2} ~ \text{d}\xi}{\sqrt{a^2 - \xi^2}} \\
&=&  -\frac{2xa}{r^2}\left(1 + \nu ~ \frac{y^2}{r^2}\right)\left[\text{K}\left(\rho\right) - \text{E}\left(\rho \right)\right] + 2\nu ~ \frac{xa^3}{3r^4}\left(3 - \frac{4x^2}{r^2}\right)\left[2\left(\text{E}\left(\rho \right) - \left(1 - \rho^2 \right)\text{K}\left(\rho \right)\right) - \rho^2 \text{E}\left(\rho \right)\right],\nonumber 
\end{eqnarray}
\begin{eqnarray}
\frac{\pi \sigma_{yy}}{2\tau_0} &=& -2\nu ~ \frac{x}{r^4}\left[x^2\int_0^r \frac{\xi^2 ~ \text{d}\xi}{\sqrt{r^2 - \xi^2}\sqrt{a^2 - \xi^2}} + \left(3 - \frac{4x^2}{r^2}\right)\int_0^r \frac{\xi^2 \sqrt{r^2 - \xi^2} ~ \text{d}\xi}{\sqrt{a^2 - \xi^2}}\right] \\
&=& -2\nu ~ \frac{xa}{r^4} \left\lbrace x^2 \left[\text{K}\left(\rho \right) - \text{E}\left(\rho \right)\right] + \left(3 - \frac{4x^2}{r^2}\right)\frac{a^2}{3}\left[2\left(\text{E}\left(\rho \right) - \left(1 - \rho^2 \right)\text{K}\left(\rho \right)\right) - \rho^2 \text{E}\left(\rho \right)\right] \right\rbrace. \nonumber
\end{eqnarray}

\subsubsection{Stresses outside the Contact Area}

\begin{eqnarray}
\frac{\pi \sigma_{xy}}{2\tau_0} &=& -\frac{y}{r^2}\left(1 - 2\nu ~ \frac{x^2}{r^2}\right)\int_0^a \frac{\xi^2 ~ \text{d}\xi}{\sqrt{r^2 - \xi^2}\sqrt{a^2 - \xi^2}} + 2\nu ~ \frac{y}{r^4}\left(1 - \frac{4x^2}{r^2}\right) \int_0^a \frac{\xi^2 \sqrt{r^2 - \xi^2} ~ \text{d}\xi}{\sqrt{a^2 - \xi^2}} \nonumber \\
&=& -\frac{y}{r}\left(1 - 2\nu ~ \frac{x^2}{r^2}\right)\left[\text{K}\left(\frac{1}{\rho }\right) - \text{E}\left(\frac{1}{\rho }\right)\right] + \\
&& \quad + 2\nu ~ \frac{ya^2}{r^3}\left(1 - \frac{4x^2}{r^2}\right)\left[\text{E}\left(\frac{1}{\rho }\right) - \frac{1}{3}\left(\left(1 + \rho^2\right)\text{E}\left(\frac{1}{\rho }\right) + \left(1 - \rho^2\right)\text{K}\left(\frac{1}{\rho }\right)\right)\right],\nonumber
\end{eqnarray}
\begin{eqnarray}
\frac{\pi \sigma_{xx}}{2\tau_0} &=& -\frac{2x}{r^2}\left(1 + \nu ~ \frac{y^2}{r^2}\right)\int_0^a \frac{\xi^2 ~ \text{d}\xi}{\sqrt{r^2 - \xi^2}\sqrt{a^2 - \xi^2}} + 2\nu ~ \frac{x}{r^4}\left(3 - \frac{4x^2}{r^2}\right) \int_0^a \frac{\xi^2 \sqrt{r^2 - \xi^2} ~ \text{d}\xi}{\sqrt{a^2 - \xi^2}} \nonumber \\
&=&  -\frac{2x}{r}\left(1 + \nu ~ \frac{y^2}{r^2}\right)\left[\text{K}\left(\frac{1}{\rho }\right) - \text{E}\left(\frac{1}{\rho }\right)\right] + \\
&& \quad + 2\nu ~ \frac{xa^2}{r^3}\left(3 - \frac{4x^2}{r^2}\right)\left[\text{E}\left(\frac{1}{\rho }\right) - \frac{1}{3}\left(\left(1 + \rho^2\right)\text{E}\left(\frac{1}{\rho }\right) + \left(1 - \rho^2\right)\text{K}\left(\frac{1}{\rho }\right)\right)\right],\nonumber
\end{eqnarray}
\begin{eqnarray}
\frac{\pi \sigma_{yy}}{2\tau_0} &=& -2\nu ~ \frac{x}{r^4}\left[x^2\int_0^a \frac{\xi^2 ~ \text{d}\xi}{\sqrt{r^2 - \xi^2}\sqrt{a^2 - \xi^2}} + \left(3 - \frac{4x^2}{r^2}\right)\int_0^a \frac{\xi^2 \sqrt{r^2 - \xi^2} ~ \text{d}\xi}{\sqrt{a^2 - \xi^2}}\right] \nonumber \\
&=& -2\nu \frac{x^3}{r^3} \left[\text{K}\left(\frac{1}{\rho }\right) - \text{E}\left(\frac{1}{\rho }\right)\right] + \\
&& \quad + \left( -2\nu ~ \frac{x a^2}{r^3}\right)\left(3 - \frac{4x^2}{r^2}\right)\left[\text{E}\left(\frac{1}{\rho }\right) - \frac{1}{3}\left(\left(1 + \rho^2\right)\text{E}\left(\frac{1}{\rho }\right) + \left(1 - \rho^2\right)\text{K}\left(\frac{1}{\rho }\right)\right)\right]. \nonumber
\end{eqnarray}

\section{Conclusions}\label{sec5}

The obtained exact solutions can serve as benchmarks for numerical models, or can be used for the deeper contact-mechanical analysis of adhesive sliding contacts of soft materials. For example, it could be considered, how the normal contact solution is influenced by the "tilting" of the contact domain due to the shear traction.

\section{Acknowledgements}

This work was supported by the German Research Foundation under the project number PO 810/66-1.

\printbibliography

\end{document}